**T2MAT (text-to-materials): A universal agent for generating material structures with goal properties from a single sentence**

Zhilong Song[1,2], Shuaihua Lu[1], Qionghua Zhou[1,2,*], and Jinlan Wang[1,2,*]

[1]Key Laboratory of Quantum Materials and Devices of Ministry of Education, School of Physics, Southeast University, Nanjing 211189, China

[2] Suzhou Laboratory, Suzhou, China

Artificial Intelligence-Generated Content (AIGC)—content autonomously produced by AI systems without human intervention—has significantly boosted efficiency across various fields. However, AIGC in material science faces challenges in efficiently discovering novel materials that surpass existing databases, while simultaneously addressing the invariance and stability of crystal structures. To address these challenges, we develop T2MAT (text-to-material), a comprehensive agent processing from a user-input sentence to inverse design material structures with goal properties beyond the existing database via globally exploring chemical space, followed by an entirely automated workflow of first-principles validation. Furthermore, we propose CGTNet (Crystal Graph Transformer NETwork), a graph neural network model that captures long-range interactions, to enhance the accuracy and data utilization efficiency of property prediction and thereby strengthen the reliability of inverse design. Through these contributions, T2MAT minimizes the dependency on human expertise and significantly improves the efficiency of discovering novel, high-performance functional materials, offering a robust way toward more autonomous materials design.

## 1. Introduction

The recent advancements in Artificial Intelligence-Generated Content (AIGC) have brought about a profound revolution across various fields, enabling efficient and high-quality AI content generation at unprecedented levels[1]. Platforms such as DeepSeek (text-to-text), MidJourney (text-to-image), and Wan (text-to-video) empower users effortlessly convert textual descriptions into sophisticated text, images, and videos, significantly boosting productivity[1,2]. Yet, in the realm of material science, while data-driven machine learning (ML)[3–8] and deep generative models[9–14] have significantly promoted materials design, current ML frameworks still demand substantial intervention from human experts. This dependency creates a bottleneck in the efficiency of ML-aided materials design, highlighting the urgent need for an advanced framework capable of achieving true text-to-material capabilities.

Unlike text-to-image or text-to-video, where the generated content only needs to align with user-defined style and content, text-to-material is a much more challenging task. Traditional ML-aided material design methodologies[3–8] are constrained to optimize existing material structures, hindering the discovery of new materials with superior properties. Thus, the key of text-to-material is to go beyond existing materials by exploring the vast chemical space, thus uncovering novel material structures with exceptional, user-specific properties. Additionally, generating material crystal structures requires strict adherence to symmetry principles, such as translational and rotational invariances. Even when a structure meets these requirements, its thermodynamic and dynamic stability must be thoroughly assessed. Unlike the subjective evaluation of generated image or videos, can be simply evaluated by human eyes or inception score[15], theoretically validating the properties of proposed material structures requires rigorous validation through computational methods like DFT calculations. These complexities remain largely unresolved in contemporary frameworks that utilize large language models to generate molecules or materials from text[16,17].

In this study, we introduce T2MAT (text-to-material), a comprehensive framework for material design that addresses the aforementioned challenges. T2MAT requires only a single sentence from the user, such as "Generate a batch of material structures with a band gap between 1-2 eV", "I want materials with SLME higher than 20% and excellent thermoelectric properties" (Figure 1). From there, the system autonomously generates novel, rational and stable material structures that meet the specified criteria. The materials generation and chemical space exploration, which is central to T2MAT, is realized by integrating two complementary frameworks: SSAGEN (Stability and Symmetry-Assured GENerative framework) and MAGECS (MAterial Generation with Efficient global Chemical space Search). SASGEN enhances structure generation through a two-stage decoupling approach that ensures both stability and symmetry constraints, while MAGECS guides generative models toward global optimization of target properties. To improve property prediction accuracy and enable reliable inverse design, we develop CGTNet (Crystal Graph Transformer NETwork), which outperforms existing GNNs in terms of accuracy and data utilization efficiency [18–24]. Additionally, T2MAT features a unified interface, which integrate both current and future GNN models developed using PyTorch Geometric[25]. By integrating contrastive learning (CL) and GNN interpretability techniques, we further enhance the accuracy, data utilization efficiency, and interpretability of property prediction models. Once crystal structures are generated, we employ an automated DFT validation framework and leverage the CSLLM (Crystal Synthesis Large Language Model)[26] to evaluate the synthesizability, synthesis methods, and precursors of the generated structures. This integration effectively bridges the gap between theoretical design and practical synthesis, providing valuable insights for experimental realization. Through these contributions, T2MAT delivers a true AIGC salution for materials design, significantly reducing reliance on human expertise, and dramatically boosting the efficiency of discovering novel, high-performance functional materials.

## 2. Results and discussion

### 2.1. Architecture of T2MAT agent

T2MAT consists of three core modules (Figure 1): i) extracting materials design requirements from a user-input sentence, ii) generating a diverse set of novel material structures that meet the specified properties, and iii) an entirely automated Density Functional Theory (DFT) validation workflow.

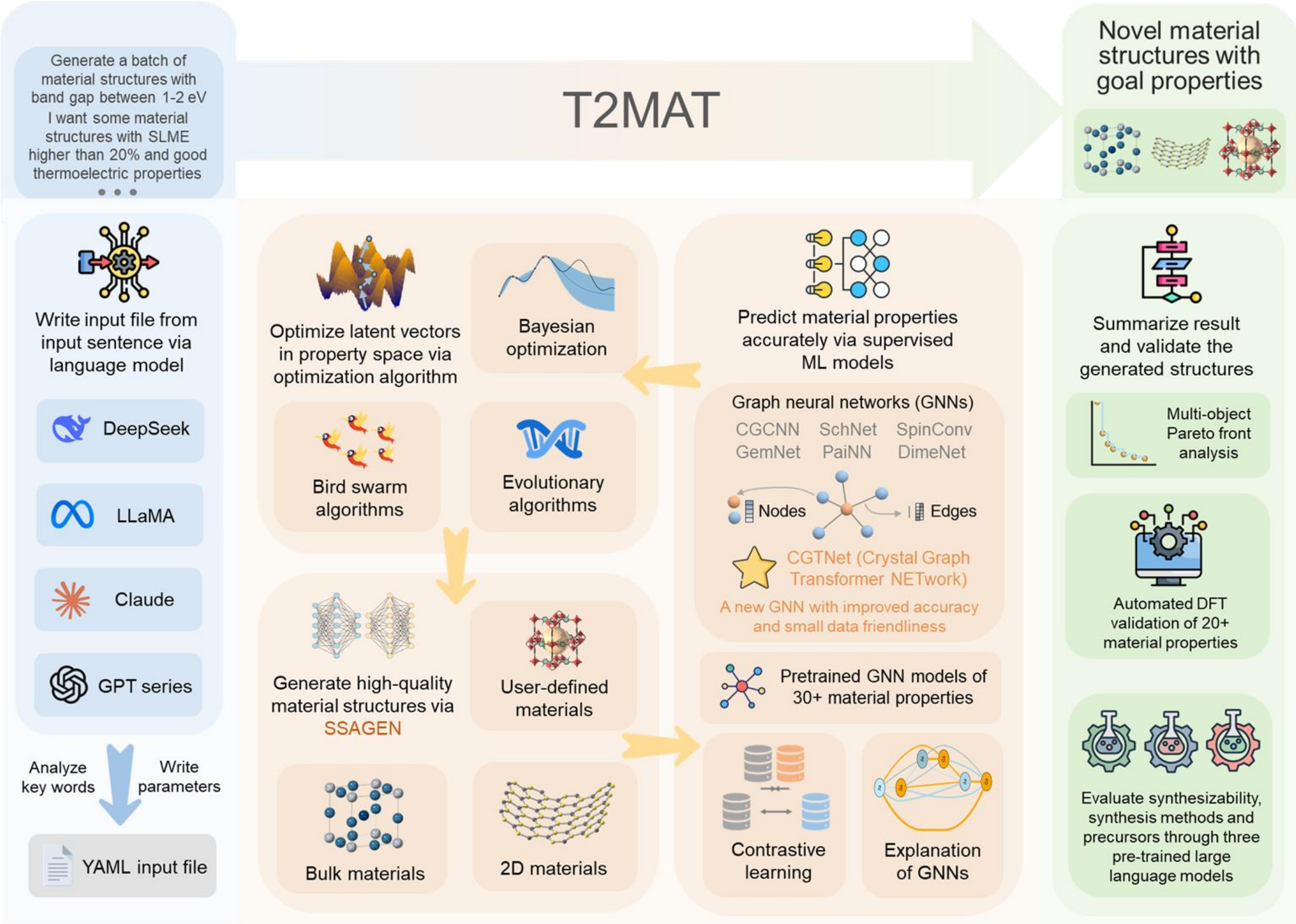

**Figure 1.** Overview of T2MAT's three primary modules: i) Capturing material design requirements from user-input (blue background); ii) Generating a variety of novel material structures based on user-specified properties (orange background); iii) Implementing a fully automated DFT validation process (green background).

In the first module, we utilize a large language model (LLM) to extract key material design requirements from user-input sentences. Currently, supported LLMs including local models such as LLaMA[27], ChatGLM[28], and Alpaca[29], as well as the online GPT[30] from OpenAI. User requirements are structured as "material property type: property value (unit)", such as, "bandgap: 1-2 eV"; "bulk modulus: 100-200 GPa". where the property value can be a range or a specific target value. For instance, if the user specifies the generation of material structures for photovoltaic applications, the property value is fixed at 1.1 eV. In addition to material properties, the system identifies the specific material design requirements by determining whether three-dimensional, two-dimensional, or MOF materials are needed, and accordingly employs specialized generative models trained on MP[31], Computational 2D Materials Database (C2DB)[32,33] and Quantum MOF (QMOF)[34,35] database, respectively. Subsequently, T2MAT creates an input file for running SASGEN and MAGECS frameworks (Figure S1 and Table S1) and defines an objective function $F$ for multi-object optimization of material properties (see details in section 3.1). To ensure the stability of generated material structures, the energy above hull is taken as an additional target property by default.

The second module aims to generate material structures and minimize the objective function $F$ of generated structures through the integrated SASGEN-MAGECS framework. The structure generation process begins with SASGEN, which transforms the conventional approach by decoupling crystal structure generation into two distinct yet complementary stages. In the first stage, SASGEN generates diverse crystal information, including lattice parameters, chemical compositions, and space groups, ensuring that fundamental crystallographic constraints are met from the onset. The second stage focuses on coordinate optimization, where SASGEN refines atomic positions to achieve thermodynamic and kinetic stability using universal machine learning potentials. This stage combines global and local optimization with symmetry and Wyckoff position constraints, and dynamically refines search spaces. Compared to conventional generative models like CDVAE, SASGEN demonstrates remarkable improvements, with a 148% enhancement in thermodynamic stability and a 180% improvement in kinetic stability, while ensuring target compositions, space groups, and lattice parameters are inherently satisfied. Once the structures are generated, their properties are predicted by advanced GNNs. T2MAT supports seven GNNs models for property prediction, including CGCNN[18], SchNet[20], SpinConv[21], GemNet[22], PaiNN[23], DimeNet++[19] and CGTNet. Since the optimal GNN may vary depending on the property being predicted, the model type is adjustable within the T2MAT input files. To expand the range of predicted properties, we have collected extensive databases linking material structures to material properties. We have trained 33 different GNNs to predict across various properties including optics, stability, mechanics, optoelectronics, magnetism, topology, thermoelectrics, piezoelectrics, dielectrics, and superconductivity. The performances of these models are summarized in Table S1. In addition,, optimization algorithms are employed to guide the SASGEN framework toward generating material structures that meet the specified properties, while efficiently exploring the immense chemical space. The MAGECS component provides outstanding capability to direct generative models toward global optimization of target properties.

The final module is designed to streamline the result analysis of MAGECS and automate the DFT validation of generated structures, eliminating the need for labor-intensive manual evaluation. After generating a large number of potential structures (e.g., $10^6$), we first visualize the variation in the mean/best value of each target property across the MAGECS optimization steps, demonstrating the effectiveness of T2MAT. The elemental and compound distributions of these structures are then analyzed to identify key factors that influence the achievement of superior target properties. Subsequently, the top N (default set to 1000) structures with optimal properties are selected and subjected to automated DFT validations (see details in Section 3.2). Finally, we leverage our CSLLM (Crystal Synthesis Large Language Model)[26] framework to provide comprehensive synthesis prediction for the validated structures. The CSLLM framework utilizes large language models to predict crystal synthesizability with 98.6% accuracy, synthesis methods with 91.2% accuracy, and synthesis precursors with an 80.2% success rate. This integration provides a fully automated pipeline from structure generation to experimental realization.

### 2.2. Architecture and performance of CGTNet

After building the T2MAT agent, we seek to improve the reliability of inverse design by improving the accuracy and data utilization efficiency of the property prediction model. The Message-passing graph neural networks (MPNNs) such as CGCNN, SchNet, GemNet, PaiNN, SpinConv, and DimeNet++ have demonstrated exceptional performance in predicting various material properties. MPNNs represent atoms as nodes and bonds as edges, with features such as atomic distances and angles encoded as graph features. Through a series of convolutional and pooling layers, MPNNs perform message passing, where information is exchanged between neighboring atoms within a specified cutoff radius.While this localized approach enables the extraction of detailed structural and chemical information, it inherently limits the model's ability to account for long-range interactions, which are often crucial for accurate material property predictions.

When it comes to dealing with long-range interactions, the self-attention mechanism in transformers[36] empowers language models to dynamically focus on different parts of a sequence (sentence), efficiently capturing long-range dependencies[37]. Therefore, we incorporated self-attention layers into CGTNet for comprehensive graph feature extraction (Figure 2). This ensures that interactions of every pair of atoms are taken into account and the interaction strengths are adaptively learned during training. Notably, no layer is adopted before the self-attention layers, which allows our model to deal with the graph features without losing information.

The extracted information for constructing graph features includes the atoms, the distance between two atoms, and the angles between three atoms, where the symmetry and invariance of the crystal structure are fully considered (see details in supplementary note 1 and 2). The angles between atoms are essential to capture the geometric properties of material structures[19,22]. GNNs like DimeNet and GemNet represent the angle information using spherical Bessel functions, which are complicated and computationally intensive[19,22]. CGTNet adopts a more concise and efficient approach to encode angle information, ensuring that all crystal symmetries are preserved (see supplementary note 1 and Figure S2-3). Then, the edge features are constructed from Gaussian smeared atomic distances and angles after passing through a fully connected (FC) layer, and then directly added to the key and value vectors in the self-attention layers (Figure 2). Atom features are derived from both atom type and specific properties, as detailed in Table S2. To encompass interactions with neighboring atoms in atom features, we have also integrated the many-body tensor representation (MBTR)[38].

However, this inclusion only marginally enhances prediction accuracy by 2% in the prediction of SLME (supplementary note 3) and extends the training duration. As a result, MBTR atom features serve as an optional hyperparameter in CGTNet and are not activated by default.

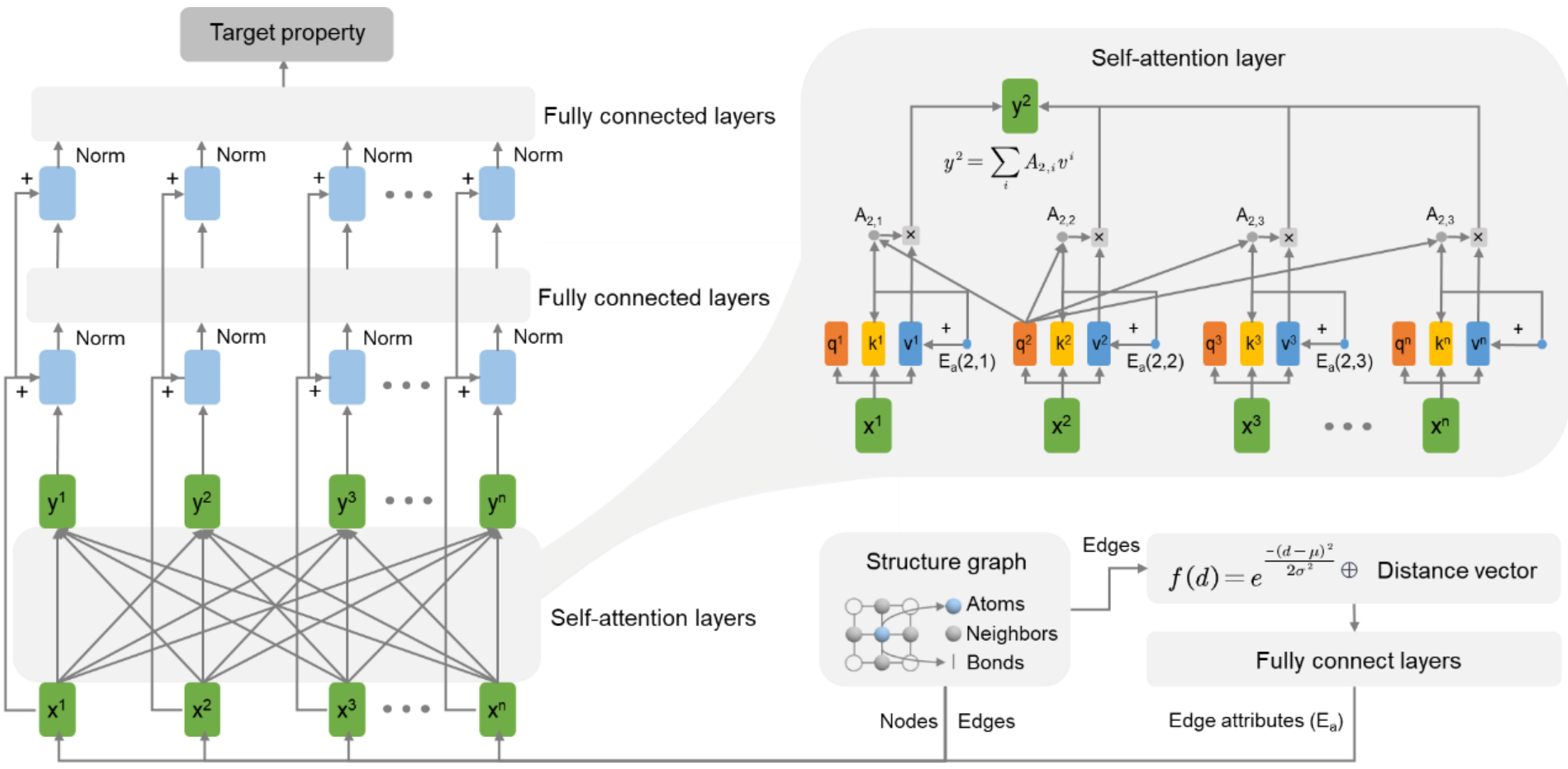

**Figure 2.** Architecture of CGTNet. The self-attention layers are designed to capture dynamically and efficiently capture long-range interaction. Graph features are constructed from atom features, interatomic distances, and angles. Edge features, processed through a fully connected layer, are integrated into the self-attention mechanism for enhanced information exchange.

To rigorously assess the accuracy and data utilization efficiency of CGTNet, we compared its performance against five other high-performance GNNs previously proposed—CGCNN, SchNet, GemNet, PaiNN, and DimeNet++—on six distinct properties, using 25%, 50%, 75%, and 100% of the training data. The properties include bulk modulus, electrical conductivity, Heyd-Scuseria-Ernzerhof (HSE) band gap, Shockley-Queisser limit for maximum efficiency (SLME)[39], maximum phonon spectrum frequency, and exfoliation energy, representing three data scales: $2\times10^4$, $10^4$, and $10^3$ samples. As shown in Figure 2, CGTNet consistently achieved the lowest the Mean Absolute Error (MAE) across the properties, outperforming all other GNN models when trained on the full dataset, with the exception of electronic conductivity. Furthermore, CGTNet exhibited remarkable data utilization efficiency. For datasets such as HSE band gap, bulk modulus, maximum phonon spectrum frequency, SLME, and exfoliation energy, CGTNet outperformed or matched the best results from the other models with just 75% of the training data. Remarkably, for SLME, CGTNet achieved 98% of the best performance using merely 50% of the training data. These results highlight CGTNet's superior predictive accuracy and data utilization efficiency, making it the default model for property prediction within T2MAT.

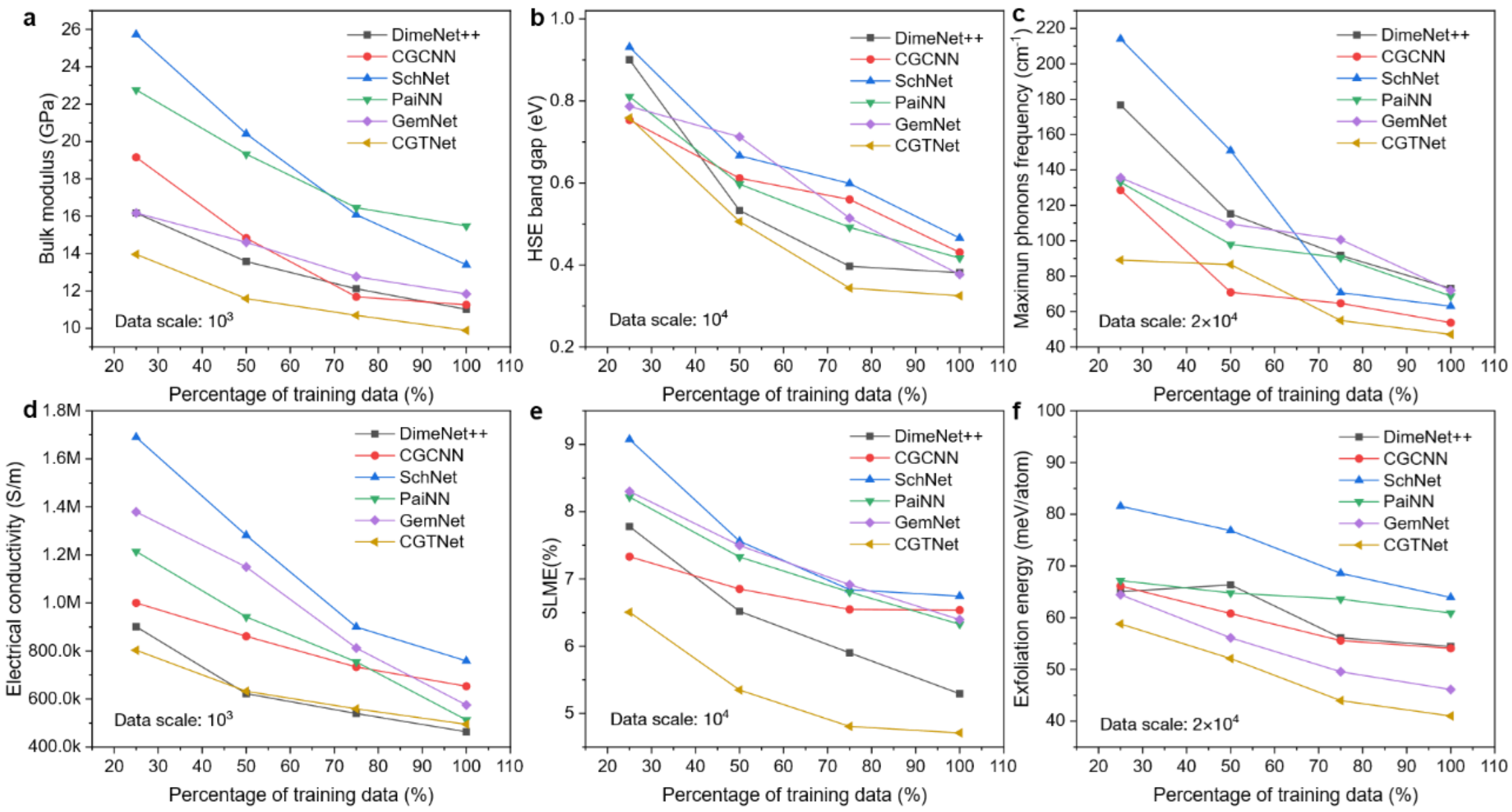

**Figure 3.** Performance of CGTNet. MAE of DimeNet++, CGCNN, SchNet, PaiNN, GemNet, and CGTNet on testing data using 25%,50%,75% and 100 % training data for the prediction of **(a)** bulk modulus, **(b)** HSE band gap, **(c)** maximum phonon spectrum frequency, **(d)** electrical conductivity, **(e)** SLME, and **(f)** exfoliation energy.

## 2.3. Integration of contrastive learning and GNN explanation

To further improve the accuracy and data utilization efficiency of GNN models, we explored the use of CL, a technique that has been successfully applied to enhance the prediction of phonon and electronic density of states[40], as well as the ground state energy in the time-dependent Schrödinger equation[41]. In the context of material science, CL trains models to differentiate between similar and dissimilar material structures through a self-supervised approach. This process creates positive pairs (two augmentations of the same material structure) and negative pairs (augmentations of different material structures). The objective is to minimize the distance between positive pairs and maximize the distance between negative pairs in the feature space. For example, given two encoded representations of material structures, $z_i$ and $z_j$, the contrastive loss function seeks to minimize $d(z_i, z_j)$ when both represent the same type of structure or property and to maximize $d$ ($z_i$, $z_j$) when they represent different structures or properties. To achieve this, the Information Noise Contrastive Estimation (InfoNCE) loss function is often employed (see details in the method section).

To quickly assess whether CL could enhance the performance of GNNs in T2MAT, we calculated the InfoNCE loss on testing data for SLME and bulk modules using 25%, 50%, 75%, and 100% of the training data. As shown in Figure 4a and b, the InfoNCE loss linearly correlates with the MAE of testing data, with $R^2$ values of 0.965 and 0.988 for SLME and bulk modules, respectively. This indicates that the predictive accuracy of the GNN improves as the model effectively minimizes distances between positive structures and maximizes those between negative pairs, resulting in a low InfoNCE loss. Furthermore, we calculated the InfoNCE loss for 17 different properties (Figure 3c), which also show a linear correlation ($R^2$ = 0.740) with their $R^2$ values on the testing data. Therefore, it is highly possible that implementing CL can further reduce the InfoNCE loss on testing data, leading to improved predictive accuracy for GNN in T2MAT.

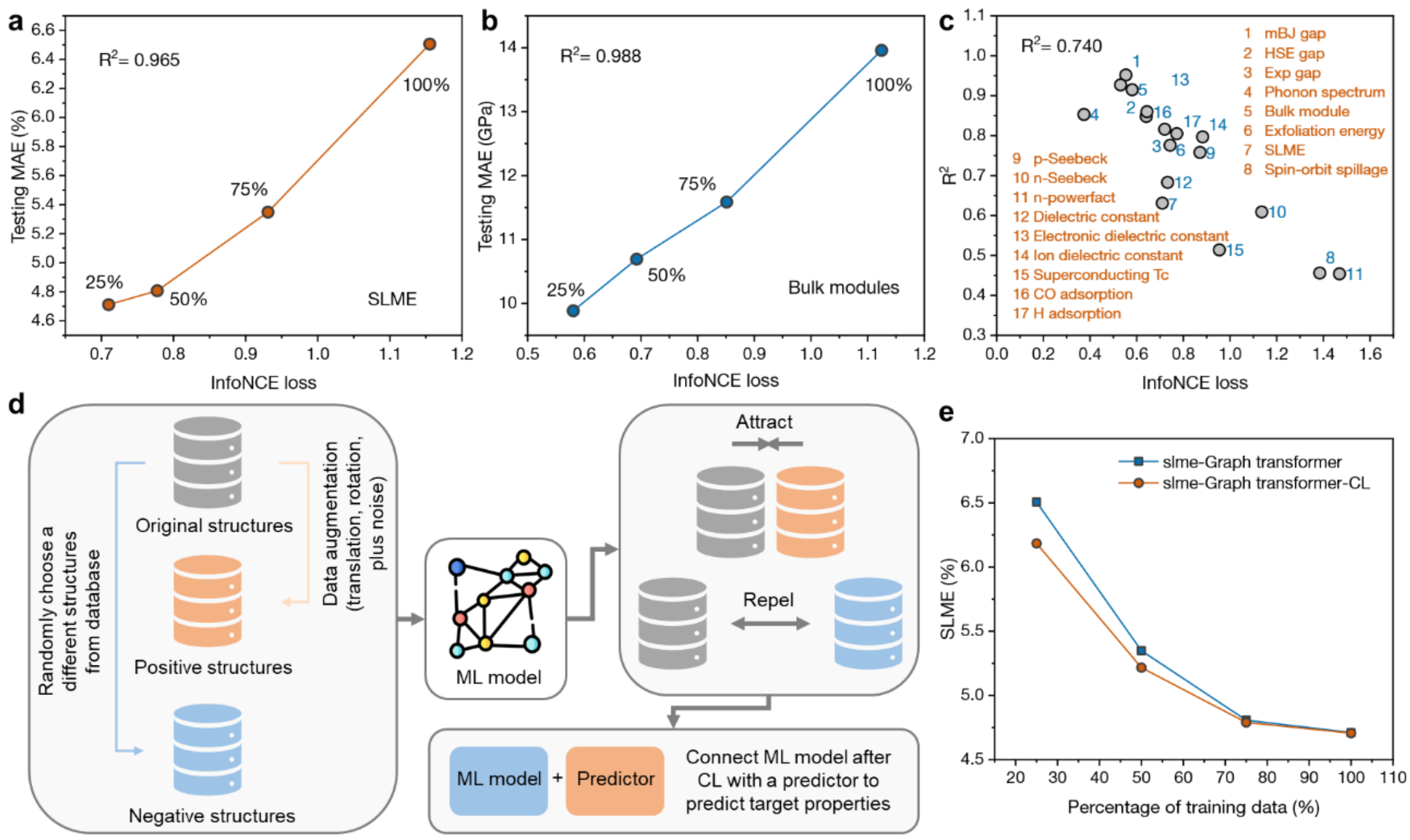

**Figure 4.** Necessity, architecture and performance of contrastive learning. **(a-c)** Linear relationship between the InfoNCE loss function and the predictive performance on testing data. **(d)** Architecture of contrastive learning for minimizing distances between positive structures and maximize those between negative ones. **(e)** Performance comparison of CGTNet

with and without contrastive learning.

We have incorporated CL into the T2MAT agent and made it compatible with all GNNs models (Figure 4d). Users can also train the system with their own models and material data. To train the CL model, we extracted a total of 896,718 material structures from the Materials Project (MP)[31], OQMD[42], and JARVIS[43] databases. Positive pairs were randomly generated from this dataset, while negative pairs were constructed by randomly translating, rotating, and adding noise to these material structures. The CL model was trained using the InfoNCE loss function and integrated with a predictor, consisting of several fully connected layers for property prediction. We then compared the performance of CGTNet with and without CL in predicting of SLME. As shown in Figure 4e, incorporating CL further improved both the accuracy and data utilization efficiency of CGTNet, especially when using smaller datasets, demonstrating the clear benefits of CL in enhancing predictive performance.

Despite their high predictive accuracy, GNNs are often considered black-box models, lacking interpretability, and failing to provide insights into the physical and chemical significance behind their prediction. To address this limitation, we employed the GNN Explainer[44], a tool specifically designed to demystify the decision-making processes of GNNs. The GNN Explainer identifies a compact subgraph and corresponding node features that are pivotal for a GNN's decision for a given node. To achieve this, it introduces masks over the graph's edges and node features, effectively highlighting the importance of each component. Through an optimization process, these masks are iteratively refined to ensure the predictions of the masked graph closely match those of the original graph. Meanwhile, a regularization term is used to promotes sparsity in the masks, retaining only the most influential edges and node features, *i.e*, the key atoms and bonds that drive the material structure's properties. This approach provides a clearer understanding of the underlying factors influencing the GNN's predictions.

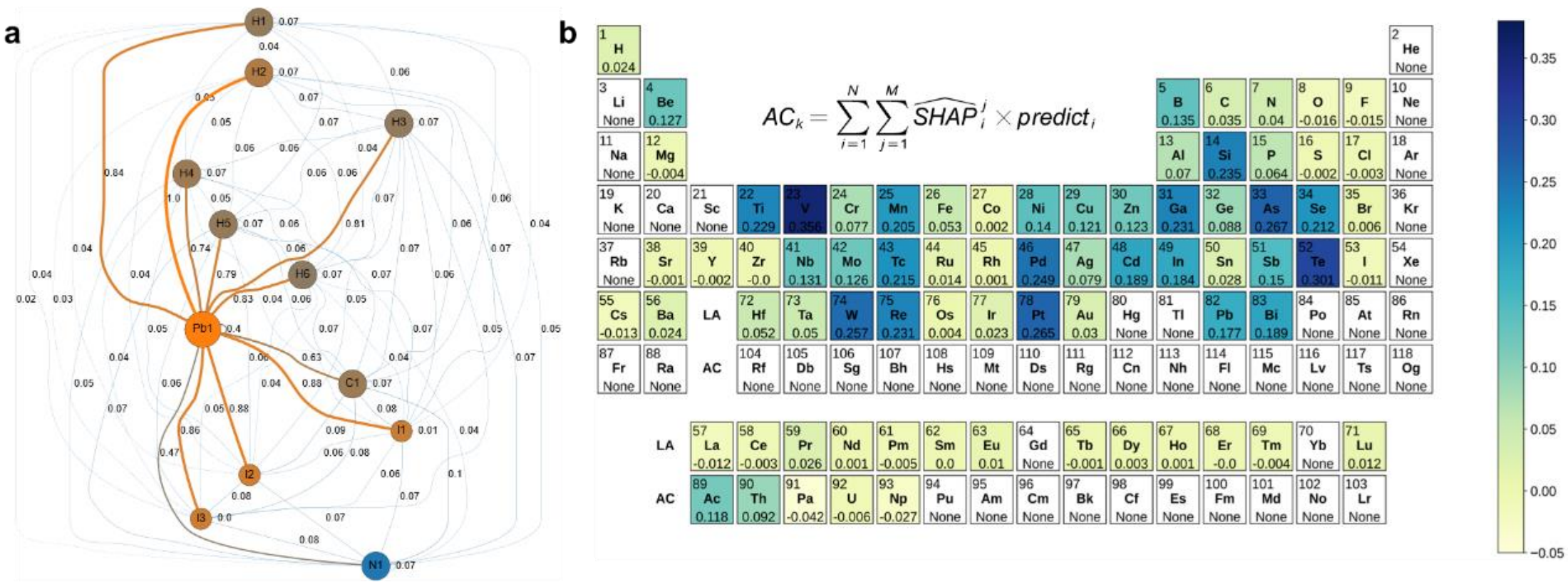

**Figure 5.** Example of the GNN explainer for interpreting GNNs in T2MAT. **(a)**Visualization of the node and edge importance in predicting the SLME of $MAPbI_3$. **(b)** Contribution of individual elements in improving the SLME, which is summarized based on 9770 GNN-based SLME predictions. The bluer an element's color, the greater its positive contribution to the SLME.

The GNN explainer is fully integrated into T2MAT for all supported GNNs. For instance, when using the CGTNet model to predict SLME, T2MAT provides detailed explanations for the model's predictions on each material structure. As illustrated in Figure 5a, we present the explanation results for the well-known photovoltaic material $MAPbI_3$[45,46], where CGTNet predicts an SLME of 21.90%. Notably, during the prediction process, CGTNet identifies the most influential nodes as lead (Pb) and iodine (I), and the most significant edge as the Pb-I bond length. This aligns with the underlying physics of $MAPbI_3$, where its CBM and VBM are predominantly contributed by the 3d orbitals of Pb and the 4s orbitals of I[46]. Consequently, the characteristics of Pb and I atoms, along with the interactions between Pb and I atoms, play a pivotal role in determining light absorption and photovoltaic efficiency.

While the GNN explainer identifies the importance of nodes and edges within the materials graph for property prediction, it does not distinguish whether these features have a positive or negative influence on the predicted properties. To address this gap and gain a deeper understanding of feature contributions, we integrated SHAP (SHapley Additive exPlanations)[47] analysis into the GNN Explainer (see details in supplementary note 4 and Figure S4-5). SHAP values provide a robust framework for quantifying the impact of individual features on model predictions, assigning each feature a numerical score that not only reflects its importance but also indicates whether it contribute positively or negatively to the predicted property. This additional layer of insight enhances the interpretability of the model by uncovering the specific factors driving predictions and increases the model's overall transparency. For the CGTNet model predicting SLME, we computed the product of the standardized SHAP values for all nodes (atoms) and edges (bonds) across the material structures, paired with their respective prediction values. As illustrated in Figure 5b and Figure S6, this analysis highlights the atoms and bonds contribute positively to the SLME, offering a clear view of how specific material features contribute to improving the photovoltaic efficiency of the structures.

### 2.4. The command line and GUI usage of T2MAT

To facilitate the user experience with T2MAT, we provide both a command-line interface and a graphical user interface (GUI). For users familiar with terminal commands T2MAT can be easily initiated by executing the run-t2mat command, followed by inputting materials design requirements directly in the terminal. This approach provides a streamlined, efficient way to those comfortable with command-line tools. For enhanced usability,To improve user experience, we also developed a GUI for T2MAT. Users can easily input their material design requirements into the text box located above the "Generate" button. Upon clicking "Generate," the system will automatically analyze and display the relevant material properties and their specified ranges. The interface is further optimized for convenience, allowing users to view and modify the input.yaml file by clicking the green "Show input.yaml" button, which displays the file content on the right side. While, the input.yaml file can be edited directly within the GUI, the default settingparameters usually do not require adjustment. After reviewing the inputs, users can initiate the inverse design process by simply clicking the blue "Submit" button, seamlessly triggering the material generation task.

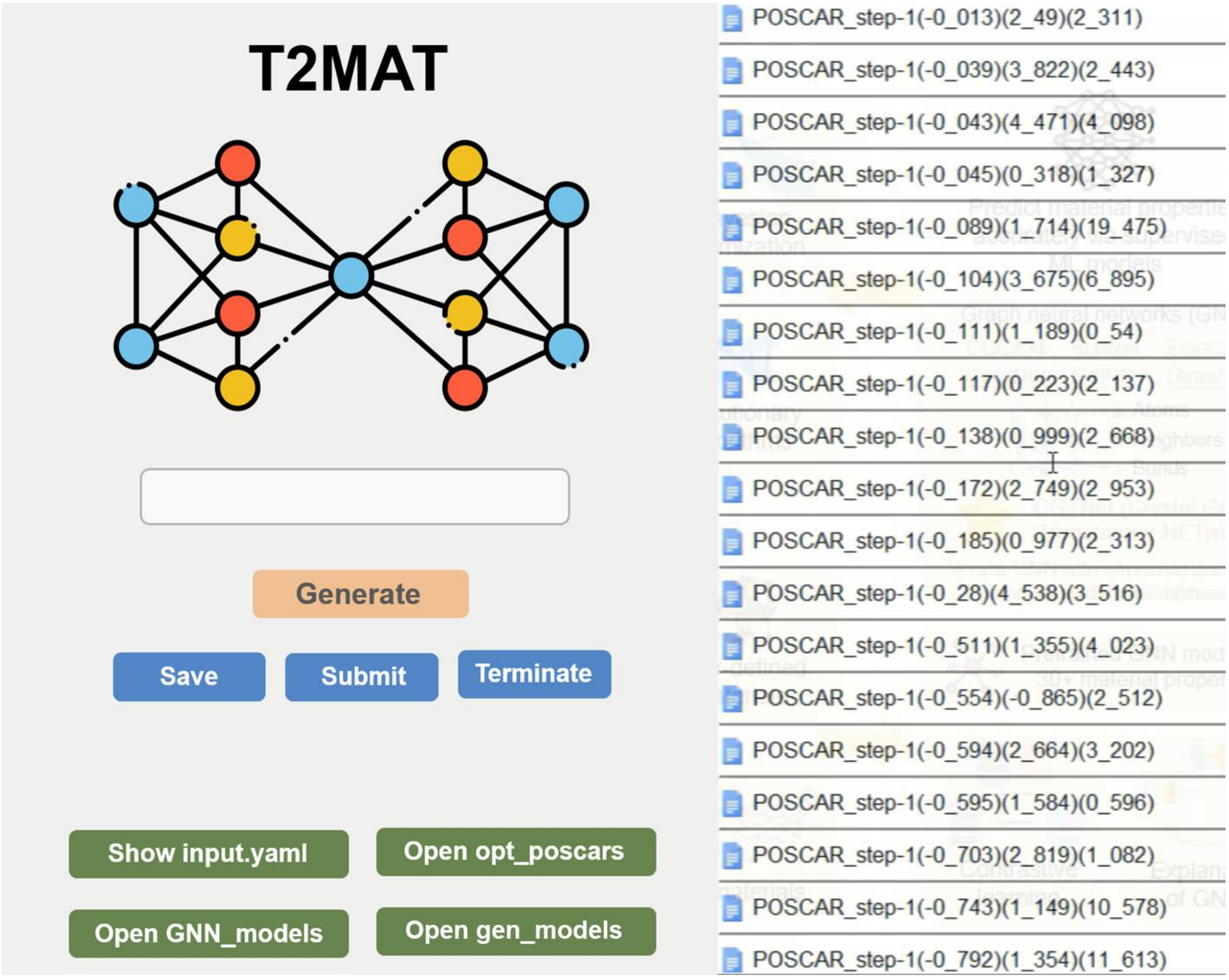

**Figure 6.** GUI of T2MAT. The interface includes the submission of material design requirements, modifications of the input file, display of available generative models, property-prediction models and generated structures.

Moreover, the GUI offers robust model management features. By selecting "Open CDVAE_models," users can navigate to a directory containing pre-trained generative models, which currently support the generation of 2D, 3D, and metal-organic framework (MOF) materials. Similarly, the "Open GNN_models" option provides access to the library of pre-trained graph neural network (GNN) models, covering 33 distinct material property prediction models. For easy viewing of generated POSCAR structure files, users can click "Show opt_poscars". Double-clicking any POSCAR file will automatically launch the VESTA program[48] , enabling seamless visualization of crystal structure visualization. These features provide users with a streamlined workflow for both model management and structure analysis.

## 3. Methods

### 3.1 Extraction of objective function for multi-object optimization in T2MAT

Utilizing LLMs, T2MAT extracts precise material design requirements and formulates the objective function $F$,

$$F = \sum_{i}^{K} C_i P_i(x_i)$$

$$P_i(x_i) = \begin{cases} 100\ I_{[R_i^1, R_i^2]}(x_i) \\ N(|x_i - R_i|) \end{cases}$$

$$I_{[R_i^1, R_i^2]}(x_i) = \begin{cases} 0\ if\ R_i^1 \leqslant x_i \leqslant R_i^2 \\ 1\ otherwise \end{cases}$$

In this function, $C_i$, $x_i$, and $R_i$ represent the weight, property value of generated structures, and user-defined range or optimal value of property $i$, respectively. $K$ is the number of required properties, and $N$ is the normalizing factor. By default, $C_i$=0 for all properties, implying an equal importance across all design criteria. However, users have the flexibility to adjust the values of $C_i$ within the input file before initiating the MAGECS process to prioritize specific properties. For instance, for a user input such as "Generate a batch of photovoltaic material structures with bulk modulus between 100-200 GPa", the objective function F [$N(|x_1 - 1.1|) + 100\ I_{[100,200]}(x_2)$], would incorporate the predicted band gap ($x_1$) and bulk modulus ($x_2$) of t

**3.2 Automatic DFT validation**

The automatic DFT validation framework (Figure S7) is implemented through the Vienna Ab initio Simulation Package (VASP)[49] and Quantum ESPRESSO (QE)[50], which are widely recognized for their robustness and accuracy in the field of computational materials science. We set default values for important parameters of automatic calculation. For example, the default values of energy cutoff, convergence criteria for electronic optimization and force calculations are set to a of 500 eV, $10^{-5}$ eV and 0.02 eV $atom^{-1}$, respectively, to balance computational efficiency and accuracy.

This framework begins with a structural relaxation module, where the input structure is optimized using a self-consistent field (SCF) calculation with the PBE functional. Then, modules are designed to calculate a broad range of material properties including stability, mechanics, optoelectronics, magnetism, topology, thermoelectrics, piezoelectrics, dielectrics, and superconductivity. Each module relies on the data generated in the previous steps, creating a seamless and systematic workflow that progresses through the necessary calculations to provide a comprehensive material property profile.

## Conclusion

In summary, we have introduced T2MAT (Text-to-material), a comprehensive material AIGC agent that enables seamless material design from a single user input sentence. T2MAT automatically generate novel, rational and stable material structures tailored to user-defined properties and performs rigorous DFT validation. The materials generation and chemical space exploration are realized by integrating the SASGEN and MAGECS frameworks, ensuring both structural rationality and discovery of high-performance materials through global optimization. The property prediction component, CGTNet, a GNN designed to effectively capture long-range interactions, offering substantial improvements in accuracy and data utilization efficiency over existing models. Additionally, by incorporating contrastive learning and GNN explanation techniques, we significantly enhance both the predictive accuracy, data utilization efficiency, and interpretability of property prediction models. Furthermore, we integrated automated DFT calculation framework and the framework to evaluate synthesizability, synthesis methods, and precursors of generated structures, bridging the gap between theoretical material design and practical synthesis. With these innovations, T2MAT revolutionizes material design, dramatically increasing efficiency and minimizing dependence on human expertise, and paving the way towards the AIGC-driven discovery of novel functional materials.

**Acknowledgements**

This work was supported by the National Key Research and Development Program of China (grant 2022YFA1503103, 2021YFA1200703), the Natural Science Foundation of China (grant 22033002, 9226111), and the Basic Research Program of Jiangsu Province. We thank the National Supercomputing Center of Tianjin and the Big Data Computing Center of Southeast University for providing the facility support on the calculations.